\documentclass[aps,twocolumn,preprintnumbers,showpacs,showkeys,nofootinbib%
]{revtex4}
\usepackage{epsfig}
\usepackage{amssymb,amsmath,amsfonts,amsthm,graphicx,psfrag}
\graphicspath{{./Figures/}}                 
\newcommand{\noi}{\noindent}
\newcommand{\beq}{\begin{equation}}
\newcommand{\eeq}{\end{equation}}
\newcommand{\bea}{\begin{eqnarray}}
\newcommand{\eea}{\end{eqnarray}}
\newcommand{\Fig}[1]{Fig.~\ref{#1}}
\newcommand{\Tab}[1]{Table~\ref{#1}}
\newcommand{\Sec}[1]{Section~\ref{#1}}
\newcommand{\Eq}[1]{Eq.~(\ref{#1})}
\newcommand{\caa}{{\cal A}}

\newcommand{\tr}{\operatorname{Tr}}

\newcommand{\bc}{{\it bc~}}
\newcommand{\fc}{{\it fc~}}
\begin{document}
\preprint{HU-EP-08/59, ITEP-LAT/2008-22}

\title{Infrared behavior and Gribov ambiguity  \\
in $SU(2)$ lattice gauge theory}

\author{V.~G.~Bornyakov}
\affiliation{Institute for High Energy Physics , 142281, Protvino, Russia \\
and Institute of Theoretical and Experimental Physics, 117259 Moscow, Russia}

\author{V.~K.~Mitrjushkin}
\affiliation{Joint Institute for Nuclear Research, 141980 Dubna, Russia \\
and Institute of Theoretical and Experimental Physics, 117259 Moscow, Russia}

\author{M.~M\"uller--Preussker}
\affiliation{Humboldt Universit\"at zu Berlin, Institut f\"ur Physik, 
Newton-Str. 15, 12489 Berlin, Germany}

\date{December 15, 2008}

\begin{abstract}

For $SU(2)$ lattice gauge theory we study numerically the infrared
behavior of the Landau gauge ghost and gluon propagators with the special
accent on the Gribov copy dependence. Applying a very efficient gauge
fixing procedure and generating up to 80 gauge copies we find that the
Gribov copy effect for both propagators is essential in the infrared.  In
particular, our {\it best copy} dressing function of the ghost propagator
approaches a {\it plateau} in the infrared, while for the random {\it
first copy} it still grows. Our {\it best copy} zero-momentum gluon
propagator shows a tendency to {\it decrease} with growing lattice size 
which excludes singular solutions. Our results look compatible with the 
so-called {\it decoupling solution} with a non-singular gluon propagator. 
However, we do not yet consider the Gribov copy problem to be finally 
resolved.

\end{abstract}

\keywords{Lattice gauge theory, gluon propagator, ghost propagator,%
Gribov problem, simulated annealing}

\pacs{11.15.Ha, 12.38.Gc, 12.38.Aw}

\maketitle

\section{Introduction}
\label{sec:introduction}

The lattice study of the gluon and ghost propagators in Landau
gauge has a long history.  It was started by Mandula and Ogilvie
\cite{Mandula:1987rh} and continued in many papers (for review see,
e.g., \cite{Giusti:2001xf,Cucchieri:2007md} and references therein).
One of the main goals of such studies was to clarify the infrared (IR)
asymptotics of the propagators and of the running coupling which can
be determined through these propagators. The hope always was that {\it
ab-initio} lattice results will give support to or discriminate between
various theoretical predictions for the IR behavior obtained with
continuum methods, in particular within the Dyson-Schwinger approach.
One prediction was definitely overturned: lattice results showed that
the gluon propagator is not divergent in the infrared.

At the same time it has been found that the lattice approach has its own
difficulties when applied to such studies. One of them is that to reach
small momenta necessary to study the IR limit one has to go to
huge lattices which makes the numerical simulations formidable. Another,
less apparent, but not less difficult problem is the problem of Gribov
copies. Although for many years it was believed that the effect of Gribov
copies on both gluon and ghost propagators was weak and could be considered
just as a noise in the scaling region
\cite{Cucchieri:1997dx,Giusti:2001xf} it has been found
first for the ghost propagator \cite{Bakeev:2003rr} and quite recently
for the gluon propagator \cite{Bogolubsky:2005wf} that these effects
are in fact quite strong. The presence of these effects makes the task of
lattice computations of the field propagators in the IR region
even more difficult.

These difficulties of the lattice approach has made it impossible so
far to obtain results which could confirm or disprove the existent
confinement scenarios proposed by Gribov~\cite{Gribov:1977wm}
and Zwanziger~\cite{Zwanziger:1993dh} on one hand and by
Kugo-Ojima~\cite{Kugo:1979gm} on the other. The Gribov-Zwanziger scenario
predicts that the gluon propagator is IR vanishing, while the Kugo-Ojima
criterion of confinement predicts the ghost dressing function to be
IR divergent.

In recent years the interest to the lattice results for the field
propagators in the IR region has been revived.  This interest was
stimulated also by the practical progress achieved over the years within
the Dyson-Schwinger (DS) approach as pursued by Alkofer, von Smekal
and others (for an intermediate review see~\cite{Alkofer:2000wg}),
and more recently with the help of functional renormalization group
(FRG) equations  ~\cite{Wetterich:1992yh,Fischer:2006vf}. Infrared QCD
has been also investigated using the stochastic quantization method
\cite{Zwanziger:2001kw,Zwanziger:2002ia}, as well as with effective
actions~\cite{Dudal:2005na,Dudal:2007cw}.

In this paper we continue our lattice study of the influence of Gribov
copies on the (minimal) Landau gauge $SU(2)$ gluon and ghost propagators
in the IR region by applying global $Z(2)$  flip transformations in
combination with an effective optimization algorithm, the so-called
{\it simulated annealing} (SA). The flip transformation was introduced
in \cite{Bogolubsky:2005wf}. Its influence on the gluon propagator was
thoroughly studied lateron \cite{Bogolubsky:2007bw}. The $Z(2)$ flips -
equivalent to non-periodic $Z(2)$ gauge transformations - were shown to
cause rather strong effects in the IR behavior of the gluon propagator. In
\cite{Bogolubsky:2007bw} high statistics computations of the gluon
propagator were made for lattice sizes varying from $1.8$~fm to $6.5$~fm
at one fixed bare lattice coupling $\beta=4/g_0^2=2.20$. The latter was
chosen in order to reach reasonably large physical volumes (and thus
small momenta) on comparatively moderate lattice sizes up to $32^4$. It
turned out that due to better gauge fixing finite-volume effects, usually
strong at minimal momenta, became largely suppressed. Furthermore,
it has been observed that at momenta $p \sim 270$~Mev the gluon
propagator seems to have a turning point leaving open the possibility
for a vanishing gluon propagator in the IR limit $p^2 \to 0$.  Here we
continue this investigation enlarging the lattice up to $40^4$ at the
same $\beta$-value corresponding to a volume of $(8.4 \mathrm{fm})^4$
and extending the studies to the ghost propagator, too. We systematically
search for Gribov copies by combining all $2^4= 16$ $Z(2)$ Polyakov loop
sectors for all Euclidean directions into one gauge orbit.

The main motivation for this computation is triggered by the puzzle posed
by the above mentioned continuum approaches.  Different
kinds of solutions with a quite different IR behavior of the gluon and
ghost propagators have been reported by different groups. 
The power-like solution with relation between gluon $\kappa_D$ and ghost 
$\kappa_G > 0$ exponents $\kappa_D = -2\kappa_G$  was called recently 
a {\it scaling solution} \cite{Fischer:2008uz}. 
This solution \cite{vonSmekal:1997is,vonSmekal:1997vx,Lerche:2002ep,
Zwanziger:2001kw,Fischer:2006vf} allows the gluon propagator
to vanish and the ghost dressing function diverge in the IR limit
in one-to-one correspondence with both the Gribov-Zwanziger
scenario \cite{Gribov:1977wm,Zwanziger:1991gz} and the
Kugo-Ojima criterion \cite{Ojima:1978hy,Kugo:1979gm} for confinement.
On the contrary, the so-called {\it decoupling solutions}
\cite{Boucaud:2007va,Dudal:2007cw,Aguilar:2008xm,Boucaud:2008ky}
provide an IR finite or weakly divergent gluon propagator and a finite 
ghost dressing function leading to a running coupling vanishing in the 
infrared. For recent discussions of the present status of research see, e.g.,
\cite{Fischer:2008uz} and further references therein\footnote{It is
worthwhile to note that the DS approach introduced originally as a method
for resummation of the perturbative series is not sensible to different
gauge copies.}.  The lattice approach based on the first-principle path
integral quantization should be able to resolve the issue.

Results for $SU(2)$ \cite{Cucchieri:2007rg} as well as
for $SU(3)$ \cite{Bogolubsky:2007ud} obtained on very large lattices and
by employing purely periodic gauge transformations seem to be in conflict
with the scaling solution and are compatible with the decoupling solution.  
That this might be not in conflict with the (appropriately
modified) Gribov-Zwanziger scenario has been recently pointed out in
\cite{Dudal:2008sp}. 

Here, by enlarging the gauge orbits with non-periodic $Z(2)$ flip gauge
transformations  and employing the SA algorithm we shall come closer
to the global extremum of the Landau gauge functional, i.e. closer to
the fundamental modular region. We find the Gribov copy dependence to be 
very strong. Still our results look rather as an argument in favor of the 
decoupling solution with a non-singular gluon propagator.
However, we do not yet consider the problem of
Gribov copies and, correspondingly, the infrared asymptotics of the
gluon propagator to be finally resolved.  

In \Sec{sec:definitions} we introduce the observables to be computed. 
In \Sec{sec:details} some details of the gauge fixing method and of the
simulation are given, whereas in \Sec{sec:results} we present our results.
Before coming to the conclusions in \Sec{sec:conclusions} we will discuss 
the dependence of our results on the number of gauge copies in 
\Sec{sec:discussion}.

\section{Gluon and ghost propagators: the definitions}
\label{sec:definitions}

For the Monte Carlo generation of ensembles of non-gauge-fixed
gauge field configurations we use the standard Wilson action,
which for the case of an $SU(2)$ gauge group is written

\bea
S & = & \beta \sum_x\sum_{\mu >\nu}
\left[ 1 -\frac{1}{2}~\tr \Bigl(U_{x\mu}U_{x+\mu;\nu}
U_{x+\nu;\mu}^{\dagger}U_{x\nu}^{\dagger} \Bigr)\right]\,, \nonumber \\
& & \beta = 4/g_0^2 \,.
\label{eq:action}
\eea

\noi Here $g_0$ is a bare coupling constant and $U_{x\mu} \in SU(2)$ are
the link variables. The latter transform as follows 
under gauge transformations $g_x$
\beq
U_{x\mu} \stackrel{g}{\mapsto} U_{x\mu}^{g}
= g_x^{\dagger} U_{x\mu} g_{x+\mu} \,,
\qquad g_x \in SU(2) \,.
\label{eq:gaugetrafo}
\eeq

\noi The standard definition~\cite{Mandula:1987rh} of the dimensionless
lattice gauge vector potential $\caa_{x+\hat{\mu}/2,\mu}$ is

\beq
\caa_{x+\hat{\mu}/2,\mu} = \frac{1}{2i}~\Bigl( U_{x\mu}-U_{x\mu}^{\dagger}\Bigr)
\equiv A_{x+\hat{\mu}/2,\mu}^a \frac{\sigma_a}{2} \,.
\label{eq:a_field}
\eeq

\noi The reader should keep in mind that the definition is not unique
which can have an essential influence on the propagator results in the
IR region, where the continuum limit is hard to control.

In lattice gauge theory the usual choice of the Landau gauge condition
is~\cite{Mandula:1987rh}

\beq
(\partial \caa)_{x} = \sum_{\mu=1}^4 \left( \caa_{x+\hat{\mu}/2;\mu}
  - \caa_{x-\hat{\mu}/2;\mu} \right)  = 0 \,,
\label{eq:diff_gaugecondition}
\eeq

\noi which is equivalent to finding an extremum of the gauge functional

\beq
F_U(g) = ~\frac{1}{4V}\sum_{x\mu}~\frac{1}{2}~\tr~U^{g}_{x\mu} \,,
\label{eq:gaugefunctional}
\eeq

\noi where $V=L^4$ is the lattice volume, with respect to gauge
transformations $g_x~$.  After replacing $U \Rightarrow U^{g}$ at the
extremum the gauge condition (\ref{eq:diff_gaugecondition}) is satisfied.
The manifold consisting of Gribov copies providing local maxima of the
functional (\ref{eq:gaugefunctional}) and a semi-positive Faddeev-Popov
operator (see below) is called {\it Gribov region} $\Omega$, while
that of the global maxima is called the {\it fundamental modular domain}
$\Lambda \subset \Omega$.  Our gauge fixing procedure is aimed to approach
this domain.

The gluon propagator $D$ and its dressing function $Z$ are then defined 
(for $p \neq 0$)  by

\bea
D_{\mu\nu}^{ab}(p)
&=& \frac{a^2}{g_0^2} 
    \langle \widetilde{A}_{\mu}^a(k) \widetilde{A}_{\nu}^b(-k) \rangle  
    \nonumber \\
&=& \left( \delta_{\mu\nu} - \frac{p_{\mu}~p_{\nu}}{p^2} \right)
    \delta^{ab} D(p)\,, 
    \nonumber \\ 
Z(p) &=& D(p)~p^2 \,,
\label{eq:gluonpropagator}
\eea

\noi where  $\widetilde{A}(k)$  represents the Fourier transform of
the gauge potentials defined by \Eq{eq:a_field} after having fixed
the gauge. The momentum $p$ is given by $p_{\mu}=(2/a) \sin{(\pi
k_{\mu}/L)}, ~~k_{\mu} \in (-L/2,L/2]$.  For $p \ne 0$, one gets from
\Eq{eq:gluonpropagator}

\beq
D(p) = \frac{1}{9} \sum_{a=1}^3 \sum_{\mu=1}^4 D^{aa}_{\mu\mu}(p) \,,
\eeq
whereas at $p = 0$ the ``zero momentum propagator'' $D(0)$ is defined as
\beq
D(0) = \frac{1}{12} \sum_{a=1}^3 \sum_{\mu=1}^4 D^{aa}_{\mu\mu}(p=0) \,.
\eeq

The lattice expression for the Landau gauge Faddeev-Popov operator 
$M^{ab} = - \partial_{\mu} D^{ab}_{\mu}$ (where $D^{ab}_{\mu}$ denotes the 
covariant derivative in the adjoint representation) for $SU(2)$ is given by 

\bea
M^{ab}_{xy}[U]  =  \sum_{\mu}~\Bigl\{ 
  \left( \bar{S}^{ab}_{x\mu} + \bar{S}^{ab}_{x-\hat{\mu};\mu}
\right)~\delta_{x;y} \Bigr. \nonumber     \\
\Bigl.   - \left( \bar{S}^{ab}_{x\mu} - \bar{A}^{ab}_{x\mu}
\right)~\delta_{y;x+\hat{\mu}} \Bigr.  \\
\Bigl.   - \left( \bar{S}^{ab}_{x-\hat{\mu};\mu}
+ \bar{A}^{ab}_{x-\hat{\mu};\mu} \right)~\delta_{y;x-\hat{\mu}} 
\Bigr\} \nonumber
\label{eq:M-form3}
\eea

\noi where

\beq
\bar{S}^{ab}_{x\mu} = \delta^{ab}~\frac{1}{2}~\tr~U_{x\mu} \,,
\quad 
\bar{A}^{ab}_{x\mu} = -\frac{1}{2}~\epsilon^{abc}~A_{x+\hat{\mu}/2;\mu}^c \,.
\label{eq:abbreviations}
\eeq

\noi From the expression (\ref{eq:M-form3}) it follows that a trivial
zero eigenvalue is always present, such that at the Gribov horizon
$\partial \Omega$ the first  non-trivial zero-eigenvalue appears.  Thus,
if the Landau gauge is properly implemented, $M[U]$ is a symmetric and
semi-positive matrix.

The ghost propagator $G^{ab}(x,y)$ is defined
as~\cite{Zwanziger:1993dh,Suman:1995zg}
\beq
G^{ab}(x,y) = \delta^{ab}~G(x-y) \equiv
\frac{1}{a^2}\Bigl<\,\left(\, M^{- 1}\,\right)^{a\,b}_{x\, y} [U]\,\Bigl> \,.
\label{ghostprop}
\eeq

\noi Note that the ghost propagator becomes translational invariant ({\it
i.e.}, dependent only on $x-y$) and diagonal in color space only in the
result of averaging over the ensemble of gauge-fixed representants of the
original Monte Carlo gauge configurations. The ghost propagator $G(p)$
in momentum space and its dressing function $J(p)$ can be written as

\bea
G(p)\, &=& \,\frac{a^2}{3 V} \sum_{x\mbox{,}\, y,\,a} 
           e^{- \frac{2 \pi i}{L} \, k \cdot (x - y)}
          \Bigl<\,\left(\, M^{- 1}\,\right)^{a\,a}_{x\, y} [U]\,\Bigl> \,, 
\nonumber \\ 
J(p) &=& G(p)~p^2\,,
\label{eq:ghostprop_in_momentumspace}
\eea

\noi where the coefficient $\frac{1}{3V}$ is taken for a full
normalization, including the indicated color average over $a=1,2,3$.
We mentioned above that $M[U]$ is symmetric and semi-positive.
In particular, it is positive-definite in the subspace orthogonal to
constant vectors. The latter are zero modes of $M[U]$.  Therefore, it
can be inverted by using a conjugate-gradient method, provided that
both the source $\psi^{a}(y)$ and the initial guess of the solution
are orthogonal to zero modes. As the source we adopted the one proposed
in \cite{Cucchieri:1997dx}

\beq
\psi^{a}(y) \,=\,\delta^{a c} \, e^{ 2 \pi i \, p \cdot y}
\qquad p \neq (0,0,0,0) \,,
\label{eq:source}
\eeq

\noi for which the condition $\sum_{y}\, \psi^{a}(y)\,=\,0$ is
automatically imposed. Choosing the source in this way allows to
save computer time since, instead of the summation over $x$ and $y$
in \Eq{eq:ghostprop_in_momentumspace}, only the scalar product of
$M^{-1}\psi$ with the source $\psi$ itself has to be evaluated. In
general, the gauge fixed configurations can be used in a more
efficient way if the inversion of $M$ is done on sources for $c =
1,2,3$ such that the (adjoint) color averaging, formally required in
\Eq{eq:ghostprop_in_momentumspace}, will be {\it explicitely} performed.

\section{Simulation details}
\label{sec:details}

We restrict ourselves to Monte Carlo (MC) simulations at $\beta=
4/g_0^2=2.20$ and use lattice field configurations for which the gluon
propagator has been already computed in \cite{Bogolubsky:2007bw}. Here we
add the computation of the ghost propagator and new data obtained on the
larger symmetric lattice with the linear size of $L=40$.  For the latter
case we generated an ensemble of 430 independent Monte Carlo lattice field
configurations. Consecutive configurations (considered as independent)
were separated by 100 sweeps, each sweep being of one local heatbath
update followed by $L/2$ microcanonical updates. In \Tab{tab:statistics}
we provide the full information about the field ensembles used throughout
this paper.

\begin{table}
\begin{center}
\mbox{
\begin{tabular}{|c|c|c|c|}   \hline
$L$ & $\# glp$ &$\# ghp$ & $N_{copy}$ \\ \hline
 8  & 200  &     & 80  \\ \hline
 12 & 200  &     & 80  \\ \hline
 16 & 240  & 60  & 24  \\ \hline
 24 & 346  & 157 & 24  \\ \hline
 32 & 247  & 118 & 40  \\ \hline
 40 & 430  & 64  & 80  \\ \hline
\end{tabular}
}
\end{center}
\caption{Lattice sizes, statistics, number of gauge copies used 
throughout this paper.  The second (third)  column gives the 
number of configurations used to compute the gluon (ghost) propagators.}
\label{tab:statistics}
\end{table}

For gauge fixing we employ the $Z(2)$ flip operation as discussed in
\cite{Bogolubsky:2007bw}. It consists in flipping all link variables
$U_{x\mu}$ attached and orthogonal to a $3d$ plane by multiplying them
with $-1$.  Such global flips are equivalent to non-periodic gauge
transformations and represent an exact symmetry of the pure gauge action
considered here. The Polyakov loops in the direction of the chosen links
and averaged over the $3d$ plane obviously change their sign. Therefore,
the flip operations combine for each lattice field configuration the $2^4$
distinct gauge orbits (or Polyakov loop sectors) of strictly periodic
gauge transformations into one larger gauge orbit.

The second ingredient of our gauge fixing procedure is the consequent
use of the simulated annealing (SA) method, which has been found
even computationally more efficient than the only use of standard
overrelaxation (OR).

The SA algorithm generates a field of gauge transformations
$~g(x)~$ by MC iterations with a statistical weight proportional to
$~\exp{(4V~F_U[g]/T)}~$. The ``temperature'' $~T~$ is a technical
parameter which is gradually decreased in order to maximize the
gauge functional $F_U[g]$.  In the beginning, $~T~$ has to be chosen
sufficiently large in order to allow traversing the configuration space
of $~g(x)~$ fields in large steps. It has been checked that an initial
value $~T_{\rm init}=1.5~$ is high enough. After each quasi-equilibrium
sweep, including both heatbath and also microcanonical updates, $~T~$ has
been decreased with equal step size until $~g(x)~$ is uniquely captured
in one basin of attraction. The criterion of success is that during the
consecutively applied OR the violation of transversality decreases in a
more or less monotonous manner for almost all applications of the compound
algorithm. This condition turned out reasonably satisfied for a final
lower temperature value $~T_{\rm final}=0.01~$~\cite{Schemel:2006yy}. The
number of temperature steps was chosen to be $1000$ for the smaller
lattice sizes and has been increased to $2000$ for the lattice size
$40^4$ included here. The finalizing OR algorithm requires a number of
iterations varying from $O(10^2)$ to $O(10^3)$. In what follows we shall
call the combined algorithm employing SA (with finalizing OR) and $Z(2)$
flips the `FSA' algorithm.

Some details of the gauge fixing procedure compared to our previous work
\cite{Bogolubsky:2007bw} have been changed. For every configuration the
Landau gauge was fixed $N_{copy}=80$ times ($5$ gauge copies for every
flip--sector), each time starting from a random gauge transformation of
the mother configuration, obtaining in this way $N_{copy}$ Landau-gauge
fixed copies. In \cite{Bogolubsky:2007bw} where smaller lattices were
simulated $N_{copy}$ was smaller: 40 (for $32^4$ lattice), 24 ($24^4$
and $16^4$). Only on very small lattices $12^4$ and $8^4$, where producing
copies was substantially cheaper, we produced 80 copies.

In order to reduce the computational effort in the finalizing OR sweeps on
the $40^4$ lattice we applied the following trick. We noticed that after
a comparably small number of OR sweeps, definitely before the convergence
criterion is reached, one can already decide which copy has a higher
maximum of the gauge functional, i.e.  one can stop the OR procedure
already when the change in the functional becomes comparably small and
further sweeps will not change the {\it order} of copies according to the
value of the maximized functional. Note, that the functional was different
from its final value only in the 8th digit and we used these values in
\Fig{fig:conv_func}.  After having selected the `best copy' (\bc)
the OR gauge fixing for this copy has to be finalized.  To be precise,
complete gauge fixing was made also on the randomly chosen `first copy'
(\fc), just for the purpose of comparison.

\vspace{5mm}

For the finalizing OR we used the standard Los Alamos type overrelaxation
with the parameter $\omega = 1.7$. For \bc and \fc the iterations have
been stopped when the following transversality condition was satisfied:

\beq
\max_{x\mbox{,}\, a} \, \Big|
\sum_{\mu=1}^4 \left( A_{x+\hat{\mu}/2;\mu}^a - A_{x-\hat{\mu}/2;\mu}^a \right)
\Big| \, < \, \epsilon_{lor} \,.
\label{eq:gaugefixstop}
\eeq

\noi We used the parameters $\epsilon_{lor}=10^{-7}$ (i.e. $10^{-14}$ for
$(\partial A)^2$).

\section{Results}
\label{sec:results}

\begin{figure}[tb]
\vspace*{-0.7cm}
\centering
\includegraphics[width=6.8cm,angle=270]{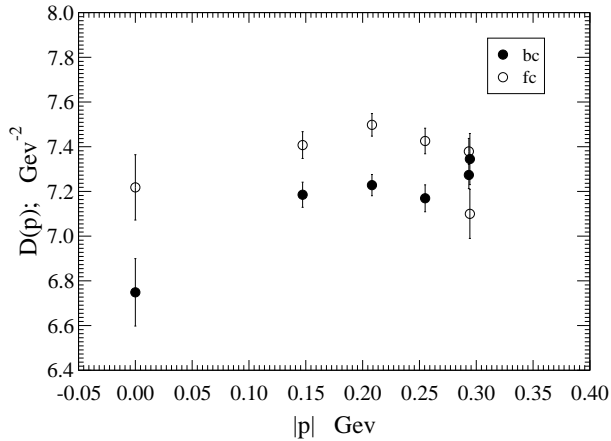}
\caption{The momentum dependence and Gribov copy sensitivity
of the gluon propagator $D(p)$ in the IR region on the $40^4$ lattice. 
Filled symbols correspond to the \bc ensemble, open symbols to the 
\fc ensemble.}
\label{fig:glp_40x40_IR}
\end{figure}

In this section we present the data for the gluon and ghost propagator. In
\Fig{fig:glp_40x40_IR} we show the new data for the gluon propagator
$D(p)$ in physical units obtained on the $40^4$ lattice at $\beta=2.20$.
We compare the \bc FSA result with the \fc SA result (the latter
without flips).  We clearly see the Gribov copy effect for the lowest
accessible momenta moving the data points to lower values for better
copies (with the larger gauge functional). The different points at $p
\sim 300 \mathrm{MeV}$ belong to different realizations of $p^2$ and
seem to indicate some violation of the hypercubic symmetry.

\begin{figure}[tb]
\centering
\includegraphics[width=6.8cm,angle=270]{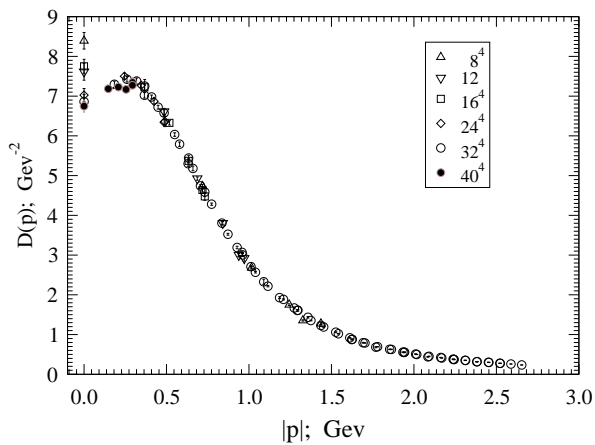}
\caption{The momentum dependence of the gluon propagator $D(p)$ on 
various lattice size. \bc results are shown throughout.}
\label{fig:glp_tot}
\end{figure}

In \Fig{fig:glp_tot} we present these new data together with the ones
obtained on smaller lattice sizes always for the FSA \bc case.

\begin{figure}[tb]
\centering
\includegraphics[width=8.5cm,angle=0]{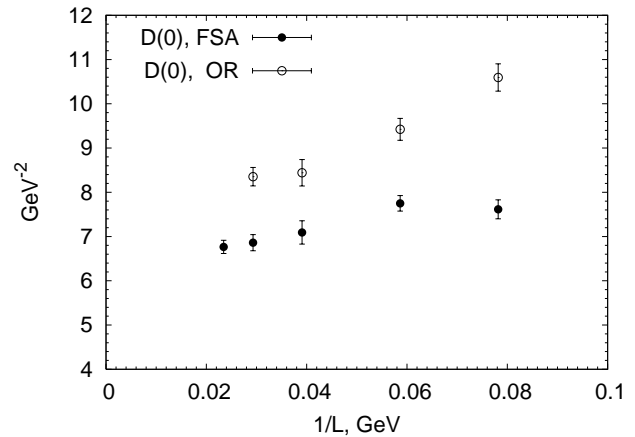}
\caption{The dependence of $D(0)$ on the lattice size. \bc FSA results 
are compared with \fc OR results (without flips).}
\label{fig:zeromom}
\end{figure}

We see that the data are nicely consistent with each other and indicate
a turnover to decreasing values towards vanishing momentum. A smooth
extrapolation to $D(0)$ becomes visible. But still there is no indication
for a vanishing gluon propagator at zero momentum for increasing volume.
This is demonstrated in \Fig{fig:zeromom}, where we show the
dependence of the zero-momentum propagator $D(0)$ as a function of the
inverse linear lattice size $1/L$. 
This behavior demonstrates a (slight) tendency to decrease, and looks
hardly consistent with $D(0)=0$ limit. One could consider it rather like
an argument in favor of the {\it decoupling} 
solution with a finite gluon propagator in the infrared.
However, one still cannot exclude that there are even more efficient
gauge fixing methods, superior to the one we use, which could make this
decreasing more drastic.

Using Ward-Slavnov-Taylor identities the authors of 
\cite{Boucaud:2007va,Boucaud:2007hy,Boucaud:2008ky} came to the conclusion
that the gluon propagator should be IR divergent, however, this divergence 
might be so weak that it could be hardly resolved on the lattice.
We believe that our results for $D(0)$ are in clear disagreement even
with a weak divergence.

\begin{figure}[tb]
\vspace*{-0.6cm}
\centering
\includegraphics[width=6.8cm,angle=270]{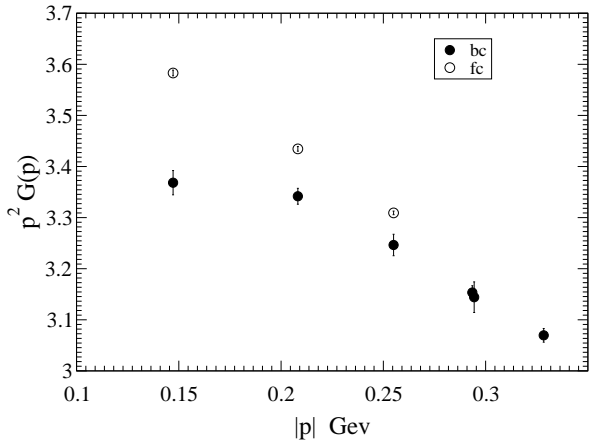}
\caption{The momentum dependence and Gribov copy sensitivity
of the ghost dressing function $J(p)=p^2\cdot G(p)$ in the IR region on the 
$40^4$ lattice. Filled symbols correspond to the \bc FSA ensemble, open 
symbols to the \fc SA ensemble.}
\label{fig:ghp_40x40_IR}
\end{figure}

Analogously to \Fig{fig:glp_40x40_IR} in \Fig{fig:ghp_40x40_IR} 
we show the ghost dressing function $J(p)$ obtained on the $40^4$ lattice.
There is a very clear Gribov copy effect changing $J(p)$ even qualitativly. Whereas the \fc SA results 
seem to support a weakly singular behavior, the \bc FSA data provide 
a {\it plateau} pointing to a finite IR value of the ghost dressing 
function, i.e. a tree-level behavior of the ghost propagator.
Our data indicate that the
plateau starts at $p \lesssim 200$ MeV.

\begin{figure}[tb]
\centering
\includegraphics[width=6.8cm,angle=270]{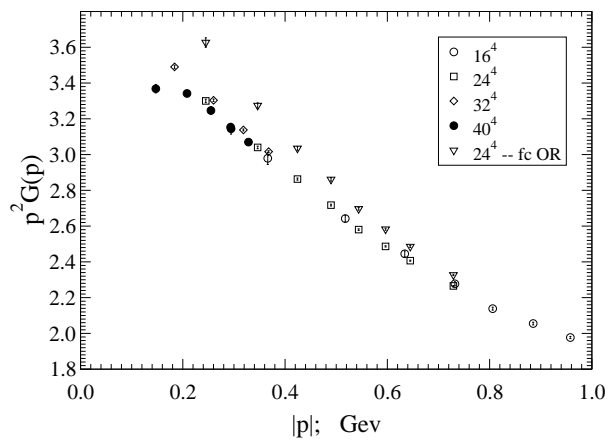}
\caption{The momentum dependence of the ghost dressing function 
$p^2\cdot G(p)$ on the various lattices. For comparison results obtained with
OR algorithm on $24^2$ lattices are also shown.}
\label{fig:ghp_tot}
\end{figure}

In \Fig{fig:ghp_tot} the ghost dressing function is shown for
lattice sizes from $16^4$ to $40^4$. We show always \bc FSA results, except
for $24^4$, where we compare also with \fc data obtained with the conventional 
OR algorithm.  The latter show an even stronger IR singular behavior than 
those data obtained with the \fc SA algorithm.

There is a clear weakening of the singularity visible additionally to a 
finite-size effect which seems to lead to an IR plateau behavior. 
Such a plateau would be consistent with the different decoupling solutions
and in contradiction with the Kugo-Ojima confinement criterion 
\cite{Ojima:1978hy,Kugo:1979gm}. 

\begin{figure}[tb]
\centering
\includegraphics[width=8.5cm,angle=0]{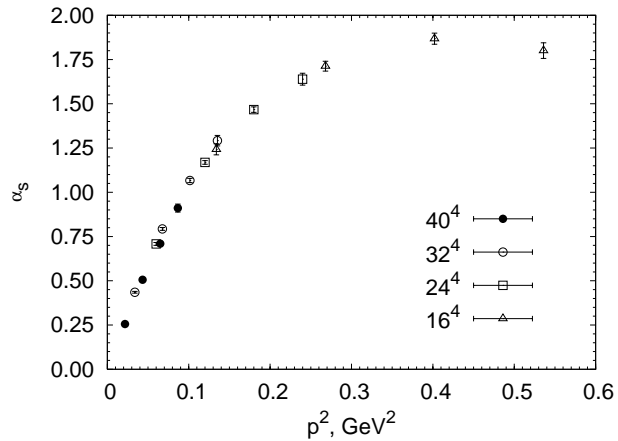}
\caption{The momentum dependence of the running coupling in the infrared region.}
\label{fig:alpha_s}
\end{figure}

In \Fig{fig:alpha_s} for the \bc FSA results obtained on lattice sizes
from $16^4$ up to $40^4$ we draw the behavior of the running coupling
related to the ghost-ghost-gluon vertex 

\beq
\alpha_s (p)= \frac{g_0^2}{4 \pi} J^2(p)~Z(p)\,
\label{eq:alpha_s}
\eeq

\noi under the assumption that the vertex function is constant as seen
in perturbation theory \cite{Taylor:1971ff} and approximately also 
in lattice simulations \cite{Cucchieri:2004sq,Ilgenfritz:2006he}.


The decrease towards $p^2=0$ is obvious. With the improved gauge fixing
the effect is even strengthened, such that an approach to an IR fixed
point as expected from the {\it scaling} DS and FRG solution seems to be 
excluded.

\section{Discussion: the quality of the gauge fixing procedure}
\label{sec:discussion}

\begin{figure}[tb]
\vspace*{-0.7cm}
\centering
\includegraphics[width=6.8cm,angle=270]{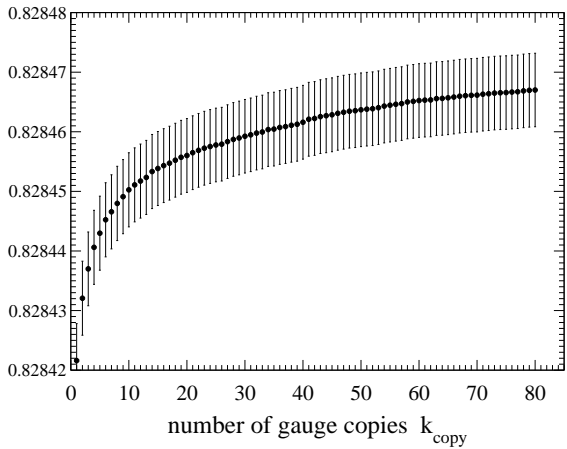}
\caption{The average value of the {\it best} gauge functional 
$\langle F_{bc}\rangle(k_{copy})$ as a function of the number
of the selected gauge copies $k_{copy}$ for $40^4$ lattice.
}
\label{fig:conv_func}
\end{figure}

In most of our simulations we have generated up to $N_{copy}=80$ gauge
copies for every thermalized configuration (up to $5$ gauge copies
for every flip-sector).  A very reasonable question is whether our results 
will change if we will further increase the number of gauge copies $N_{copy}$.
In \Fig{fig:conv_func} we show the dependence of the average \bc functional 
$\langle F_{bc} \rangle(k_{copy})$ :

\beq
\langle F_{bc}\rangle(k_{copy})  = \frac{1}{n}\sum^{n} F_{bc}(k_{copy})~,
~~k_{copy}= 1,\ldots, N_{copy}~,
\eeq

\noi where for every configuration $F_{bc}(k_{copy})$ is the 'best' (i.e.,
maximal) value of the functional $F$ found after employing $k_{copy}$
copies, and $n$ denotes the number of configurations.

One can see that this average still keeps a tendency to grow, which
could mean that one should take even more copies to reach the
global maxima. To understand it better we generated $25$ configurations
on the $40^4$ lattice with $N_{copy}=320$ (i.e., $20$ gauge copies per
sector).

\begin{figure}[tb]
\centering
\includegraphics[width=6.6cm,angle=270]{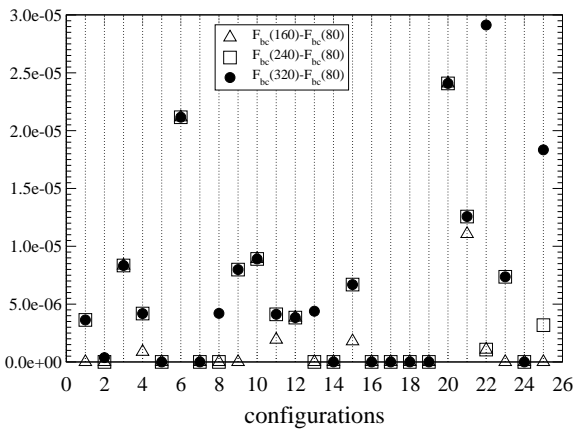}
\caption{The difference $F_{bc}(k_{copy})-F_{bc}(k_{copy}=80)$
for various numbers of the gauge copies $k_{copy}$ for $40^4$
lattice.
}
\label{fig:conv_func2}
\end{figure}

In \Fig{fig:conv_func2} we show the difference 
\beq
\Delta F(k_{copy}) =F_{bc}(k_{copy})-F_{bc}(80)
\eeq

\noi at $k_{copy}=160, 240$ and $320$ for these configurations.

For the majority of configurations this difference is rather
small. However, for about $20\%$ of configurations the difference is of
the order of $10^{-5}$, which could mean still a rather strong influence
on the values of propagators.

\begin{figure}[tb]
\centering
\includegraphics[width=8.5cm,height=5.8cm,angle=0]{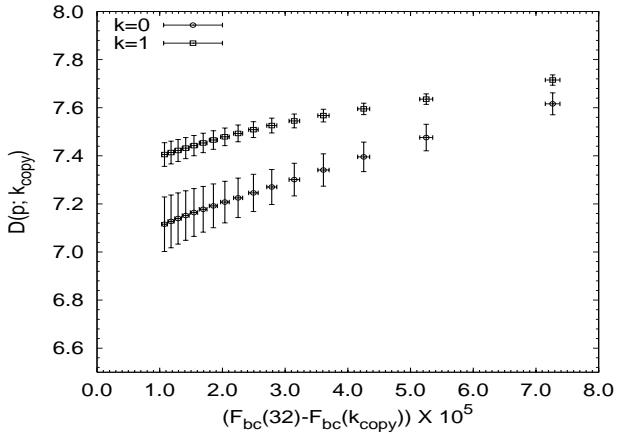}
\caption{The gluon propagator at momenta $p=0$ and $p=p_{min}$ vs 
the functional  $F_{bc}$ for $32^4$ lattice.
}
\label{fig:glp_vs_f}
\end{figure}

To demonstrate that the change in the functional of the order of $10^{-5}$
indeed might give rise to substantial change in the propagators we plot
in \Fig{fig:glp_vs_f} the gluon propagator computed on $32^4$ lattice
at two momenta, $p=0$ and $p=p_{min}$  as a function of the difference
$F_{bc}(32)-F_{bc}(k_{copy})$ for $k_{copy}=1,2,...,18$.  One can see
that the change in the functional  $F_{bc}$ in the 5th digit brings
quite substantial change in the propagator at both momenta.

These observations give an idea that there might be {\it another} even 
more efficient gauge fixing method which will be more successful in the
search of the global maximum of the functional $F$ 
(as FSA is superior with respect to standard OR).

Therefore, we cannot draw a final conclusion before spending much
more efforts into optimization of the gauge fixing procedure. 
However, we expect that both the gluon propagator as well as the ghost 
propagator will become further suppressed in the infrared, when approaching 
the fundamental modular region.

\section{Conclusions}
\label{sec:conclusions}

In this work we studied numerically the dependence of the Landau gauge
gluon and ghost propagators, as well as of the running coupling constant, 
in pure gauge $SU(2)$ lattice theory in the infrared region. The special
accent has been made on the study of the dependence of these `observables'
on the choice of Gribov copies.

The simulations have been performed using the standard Wilson action at
$\beta=2.20$ for linear lattice sizes up to $L=40$. For gauge fixing
gauge orbits  enlarged by $Z(2)$ flip operations were considered with
up to $5$ gauge copies in every flip-sector (in total, up to $80$ gauge
copies).  For $25$ thermalized configurations we  produced $20$ copies
per sector (in total, $320$ gauge copies for every configuration).
The maximization of the gauge functional was achieved by the
simulated annealing method always combined with consecutive overrelaxation
(`FSA' algorithm).

Our findings can be summarized as follows.

\vspace{2mm}
{\bf 1)} For the gluon propagator our new data for the $40^4$
lattice agree with data on the smaller lattices (up to $32^4$).
We confirm our conclusion \cite{Bogolubsky:2007bw} about the appearance
of the local maximum at a non-zero value of the momentum $p^2$ (this local
maximum was absent for lattice sizes $ \le 24^4$).

The zero-momentum gluon propagator $D(0)$ has a tendency to decrease
with growing lattice size $L$.
This observation is in  clear contradiction with the infrared
divergent gluon propagator obtained on the basis of Ward-Slavnov-Taylor 
identities.

At the time being, this behavior looks hardly consistent with a $D(0)=0$
limit at infinite $L$, and could be considered rather like an argument
in favor of the {\it decoupling} solution with a non-singular gluon propagator. 
However, we do not yet consider the problem of the infrared asymptotics
of the gluon propagator as a finally resolved (see below).

\vspace{2mm}
{\bf 2)} We calculated the ghost propagator for lattices up to $40^4$.
Our \bc dressing function $J(p)$ of the ghost propagator demonstrates
the approach to a {\it plateau} in the infrared, while the \fc dressing
function still grows (as it was in earlier calculations; see, e.g.,
\cite{Bogolubsky:2007ud,Cucchieri:2007md,Cucchieri:2008fc}).

This is a first clear indication of the lack of the IR-enhancement of
the ghost propagator.  This plateau behavior is in a clear contradiction
with the Kugo--Ogima confinement criterion. The fate of this confinement
criterion still needs a further clarification.

\vspace{2mm}
{\bf 3)} We have found that the effect of Gribov copies for both the
propagators and in the consequence for the running coupling is essential
in the infrared range $p < 1$ GeV.  Therefore, the {\it quality} of
the gauge fixing procedure in the study of gauge dependent observables
remains important.

Indeed, the FSA method provides systematically higher values of the
functional $F_U(g)$ as compared to the standard OR procedure for the
same thermalized configurations.  This means that in practice OR needs
many more random copies to explore (correspondingly, much more CPU
time to spend) to find larger values of $F_U(g)$ as compared to FSA.
This effect becomes stronger with increasing the volume.  However,
we cannot say that we have reached the fundamental modular region when
fixing the Landau gauge on larger lattices.  One cannot exclude that
there is another method superior to our FSA algorithm.  We believe that
the Gribov problem deserves even more thorough studies.

\vspace{5mm}

Maybe, there are alternative ways to resolve the problem of the 
IR asymptotics of the propagators.  We have to be aware that the lattice
method as it is normally used has still some uncertainties. First
of all the continuum limit in the infrared range is hard to
control and it depends on the proper choice of the gauge potential
$\caa_{x\mu}$. Moreover, the infrared limit is sensitive to the boundary
conditions, which normally are taken to be periodic.  Incomplete gauge
fixing in combination with these choices seems unavoidably to lead to
zero-momentum modes not sufficiently suppressed even in the thermodynamic
limit.  That the presence of zero-momentum modes can spoil the behavior
of gauge-variant propagators is well-known from the example of 4d compact
$U1)$ lattice gauge theory 
\cite{Mitrjushkin:1996fw,Bogolubsky:1999cb,Bogolubsky:1999ud}.  
Whether a BRST conformal lattice reformulation will
solve the issue as proposed in \cite{vonSmekal:2007ns,vonSmekal:2008en}
remains to be seen.

\subsection*{Acknowledgments}
This investigation has been partly supported by the Heisenberg-Landau
program of collaboration between the Bogoliubov Laboratory of Theoretical 
Physics of the Joint Institute for Nuclear Research Dubna (Russia) and 
German institutes and partly by the joint DFG-RFBR grant 436 RUS 113/866/0-1
and the RFBR-DFG grant 06-02-04014. VB and VM are supported by the 
grant for scientific schools NSh-679.2008.2. VB is supported by grants  
RFBR 07-02-00237-a and RFBR 08-02-00661-a. MMP gratefully acknowledges
useful discussions with E.-M. Ilgenfritz, A. Maas, J. Pawlowski, 
C. Fischer, O. Pene, L. von Smekal, and A. Sternbeck.


\end{document}